\documentclass[useAMS,usenatbib]{mn2e}
\usepackage{epsfig}
\usepackage{threeparttable}
\usepackage{amsmath}

\title[An alternative formation mechanism of compact IMBPs]{Magnetic braking of Ap/Bp stars: an alternative formation mechanism of compact intermediate-mass binary pulsars}
\author[W. M. Liu et al.]{Wei-Min Liu$^{1}$, Wen-Cong Chen$^{1,2}$\thanks{E-mail:
chenwc@pku.edu.cn}\\ $^1$ School of Physics, Shangqiu Normal University, Shangqiu 476000, China;\\
 $^2$ Key Laboratory of Modern Astronomy and Astrophysics (Nanjing University), Ministry of Education, Nanjing 210093, China }
\begin{document}

\date{}

\pagerange{\pageref{firstpage}--\pageref{lastpage}} \pubyear{2014}

\maketitle

\label{firstpage}

\begin{abstract}
It is difficult for the intermediate-mass X-ray binaries
(IMXBs) evolutionary channel to form intermediate-mass binary
pulsars (IMBPs) with a short orbital period (less than 3 d) via
stable mass transfer. The main reason is that the magnetic braking
mechanisms are generally thought not to work for donor stars with
a mass of greater than 1.5 $\rm M_{\odot}$ in the canonical model.
However, some intermediate-mass stars have anomalously strong
magnetic fields (about 100 -- 10000 G), i. e. so-called Ap or Bp
stars. With the coupling between the magnetic field and the
irradiation-driven wind from the surface of Ap/Bp stars, a
plausible magnetic braking mechanism should be expected. In this
work, we attempt to investigate if IMXBs with Ap/Bp stars can
produce IMBPs with a short orbital period (less than 3 d) by such
an anomalous magnetic braking mechanism. Using a stellar evolution
code, we have simulated the evolution of a large number of IMXBs
consisting of a NS and an Ap/Bp star. For the spin evolution of
the NS, we consider the accretion torque, the propeller torque,
and the spin-down torque caused by the interaction between the
magnetic field and the accretion disc. The calculated results show
that, employing anomalous magnetic braking of Ap/Bp stars, IMXBs
can evolve into compact IMBPs with short orbital periods of less
than 3 d. However, there exists significant discrepancy between
the spin periods of IMBPs in our simulated results and those
observed.
\end{abstract}


\section{Introduction}

Binary millisecond pulsars (BMSPs) consist of a millisecond radio
pulsar and a white dwarf (WD). Their observed orbital and stellar
parameters are fossils of binary stars evolution. Thus BMSPs can
be used as important probes to test stellar astrophysics and
binary evolutionary theory. According to the nature of the WD
companions, BMSPs were divided into two groups: low-mass binary
pulsars (LMBPs) and intermediate-mass binary pulsars (IMBPs).
Nowadays, it is generally accepted that most LMBPs are evolved
from neutron star (NS) low-mass X-ray binaries (LMXBs). By
Roche-lobe overflow of the progenitor of the WD, NS can be spun up
to a millisecond period via the accretion of material and angular
momentum from the donor star \citep{alpa82,radh82}. At the same
time, the mass accretion induces the NS magnetic fields decrease
to about $10^{8} - 10^{9}$ G \citep{kona97}. The final
evolutionary endpoints are LMBPs consisting of a millisecond
pulsar and a low-mass (less than 0.4\,$\rm M_{\odot}$) He WD (see
Bhattacharya \& van den Heuvel 1991; Tauris \& Savonije 1999;
Podsiadlowski et al. 2002; Tauris \& van den Heuvel 2006).

Comparing with LMBPs, IMBPs consist of a massive (greater than
0.4\,$\rm M_{\odot}$) CO or ONeMg WD and a mildly recycled pulsar
(Camilo et al. 1996, 2001; Edwards \& Bailes 2001a), in which the
spin periods and magnetic field strengths are considerably greater
than those of LMBPs \citep{li02}. Furthermore, IMBPs tend to have
a relatively short orbital-period (less than 40 d) and a
relatively large orbital eccentricity. These properties imply that
there exist different progenitor systems and evolutionary
histories between IMBPs and LMBPs. The CO WDs can evolve from
normal intermediate-mass hydrogen-rich stars on asymptotic giant
branch by burning hydrogen and helium shells around the CO core
\citep{heuv94}. Therefore, it is generally agreed that IMXBs can
evolve into IMBPs. Due to the large mass ratio (more than 1.5) of
IMXBs, the mass transfer was thought to be unstable, and
lead to a common envelope evolution \citep{pacz76,webb84,iben93}.
With the spiral-in of the NS in the envelope of the donor star, it
will be mildly recycled by a short timescale super-Eddington
accretion \citep{heuv84,stai04}. However, a NS may collapse into a
black hole during hypercritical accretion in the common envelope
phase \citep{chev93,brow95,brow01}. Subsequently, it was
discovered that, except for the donor stars with a deeply
convective envelope, IMXBs with a donor star of 2 -- 6 $\rm
M_{\odot}$ can avoid a spiral-in stage and evolve into IMBPs by
rapid mass transfer on a thermal timescale
\citep{taur00,pods02,pfah03}. Based on the isotropic re-emission
model, \cite{taur00} found that IMXBs can evolve into IMBPs with
relatively long orbital period ($P_{\rm orb} \approx$ 3 - 50 d).
Recently, the population-synthesis calculation also shows that it
is difficult to form BMSPs with $P_{\rm orb} <10$ d by the
standard binary star evolution \citep{hurl10}.

At present, there exist 9 compact IMBPs with an orbital period of
less than 3 d. Therefore, it is very interesting for the stellar
evolutionary theory to account for the origin of these compact
IMBPs. These systems could be explained if the NS survives a phase
of common envelope (CE) evolution with an intermediate-mass donor
star. The post-CE system consists of a NS and a He star (the naked
core of the donor star). Because He stars with a mass less than 2
$\rm M_{\odot}$ can evolve into CO WDs by burning their helium
shell \citep{habe86}, NS + He star binaries can form most of the
observed IMBPs by Case BB mass transfer from naked He stars
\citep{taur12b}. It is worth noting that NS + He star binaries
have an ultra-short orbital periods in the range of 0.01 -- 1.0 d
due to common envelope evolution \citep{chen11}. Recently, Chen \&
Liu (2013) have investigated the initial parameter space (He star
masses and initial orbital periods) of He star + NS binaries that
can form IMBPs. Their simulated results show that the NS + He star
evolutionary channel can explain the formation of 4 -- 5 short
orbital period IMBPs. The discrepancies between the simulation and
the observations may originate from the accretion model of the NS
and the magnetic braking scenario, which should be improved.

The aim of this work is to investigate whether the IMXBs
evolutionary channel can form compact IMBPs without evolving
through a CE-phase. The majority of intermediate-mass stars (more
than 1.5$\rm M_{\odot}$) with radiative envelopes are not thought
to experience magnetic braking. However, about 5\,\% of A/B type
stars possess anomalously strong magnetic fields (about 100 --
10000 G), and they are called Ap/Bp stars\footnote{\cite{ferr09}
argued that the upper-main sequence stars with strong magnetic
field originated from a merge of two protostars. The strong
differential rotation drove by merger process can lead to a
large-scale dynamo field. } (see also Landstreet 1982; Moss 1989;
Shorlin et al. 2002; Braithwaite \& Spruit 2004). \cite{just06}
found that magnetic braking scenarios of Ap/Bp stars can account
for the formation of compact black hole X-ray binaries with short
orbital periods of less than 1 d.

Unfortunately, the operation process of magnetic dynamo in the
early-type stars with a radiative envelope is still unclear
(Dervi\c{s}o\u{g}lu et al. 2010). In this work, solar-like
magnetic braking concepts are simply applied to intermediate-mass
stars with an anomalous magnetic field. We apply this anomalous
magnetic braking model to the evolution of IMXBs, and test whether
IMXBs can provide an alternative evolutionary channel to compact
IMBPs. This paper is structured as follows. In section 2, we will
give a detailed description for the binary evolution calculation
of IMXBs. The calculated results are presented in section 3.
Finally, we give a brief summary and discussion in section 4.

\section{Input physics}

\subsection{Stellar evolution code}
Based on an updated version of the stellar evolution code
originally developed by Eggleton (1971,1972,1973; see also Han et
al. 1994; Pols et al. 1995),  we attempt to simulate the
evolutionary sequences of IMXBs consisting of a NS (of mass
$M_{\rm NS}$) and an Ap/Bp donor star (of mass $M_{\rm d}$). The
Ap/Bp donor star is assumed to be a solar chemical composition ($X
= 0.70, Y = 0.28, Z = 0.02$), the ratio of the mixing length to
the pressure scale height and the convective overshooting
parameter are set to be 2.0 and 0 \citep{dewi02}. The stellar
OPAL opacity table is taken from \cite{chen07}, Rogers \& Iglesias
(1992), and Alexander \& Ferguson (1994). We stop the calculation
when the system evolves into a detached binary, and the donor star
evolves into a WD.

\subsection{Mass transfer and angular momentum loss}
The mechanisms driving the mass transfer in IMXBs should be the
loss of orbital angular momentum via gravitational wave radiation
and/or magnetic braking (for narrow systems with initial orbital period
of 1 -- 2 d), and the nuclear evolution of the donor
star (for relatively wide systems with initial orbital periods of greater
than 1 -- 2 d ). In our calculation, we considered three kinds of angular
momentum loss as follows.

1. Gravitational wave radiation. This angular momentum loss rate
is given by \citep{land75}
\begin{equation}
\dot{J}_{\rm GR}=-\frac{32G^{7/2}}{5c^{5}}\frac{M_{\rm
NS}^{2}M_{\rm d}^{2}M^{1/2}}{a^{7/2}},
\end{equation}
where $G$ and $c$ are the gravitational constant and the speed of
light in vacuum, respectively. $M=M_{\rm NS}+M_{\rm d}$ is the
total mass of the binary, and $a$ is the binary separation.

2. Mass loss. For IMXBs, once the donor star overflows its
Roche lobe, thermal timescale mass transfer is triggered  because the
material is transferred from the more massive donor star to the
less massive NS. The mass transfer rate $-\dot{M}_{\rm d}$ may be
greater than the Eddington accretion rate of the NS, in which
$\dot{M}_{\rm Edd}\simeq 1.5\times 10^{-8}~{\rm M_{\odot}
yr}^{-1}$ for H-rich accreted material. Therefore, the mass-loss
rate of the system can be written as \textbf{$|\dot{M}| = |\dot{M}_{\rm
d}| - |\dot{M}_{\rm Edd}|$}. For the angular momentum loss accompanying the mass
loss, we assume an isotropic re-emission scenario \citep{sobe97},
in which the excess material is ejected in the vicinity of the NS,
and carries away the specific orbital angular momentum of the NS.
The orbital angular momentum loss rate by the isotropic
re-emission is given by
\begin{equation}
\dot{J}_{\rm IR}=\frac{\dot{M}_{\rm d}M^{2}_{\rm
d}}{M^{2}}a^{2}\Omega\beta,
\end{equation}
where $\Omega$ is the orbital angular velocity of the binary
system, $\beta$ is the fraction of material lost from the donor
star which is re-emitted from the NS {\citep{sobe97}}.

3. Magnetic braking of Ap/Bp stars.  The interaction between the
magnetic field of a donor star and its stellar winds can extract
the spin angular momentum \citep{webe67,mest87,kawa88}.Similarly
to \cite{just06}, we assume that the stellar wind is bound in the
magnetic field lines to co-rotate with the stars out to the
magnetospheric radius ($r_{\rm m}$). However, the tidal
interaction between two components would accelerate the donor star
to co-rotate with the orbital motion, and indirectly remove the
orbital angular momentum from the binary. Therefore, the loss rate
of angular momentum by magnetic braking is
\begin{equation}
\dot{J}_{\rm MB}=-\Omega r_{\rm m}^{2}\dot{M}_{\rm wind}=-(GM_{\rm
d})^{-1/4}\Omega B_{\rm d}R_{\rm d}^{13/4}\dot{M}_{\rm
wind}^{1/2},
\end{equation}
where $\dot{M}_{\rm wind}$ is the stellar wind-loss rate; $B_{\rm
d}$, and $R_{\rm d}$ denote the surface magnetic field, and the
radius of the Ap/Bp star, respectively.

Actually, Ap/Bp stars in main sequence stage are not
observed to have spun down because of weak winds. However, in a
close binary system X-rays generated by accretion on to the NS can
drive a strong stellar wind from the Ap/Bp star \citep{rude89a,tava93}.
Therefore, the anomalous magnetic braking of Ap/Bp
star provides an efficient mechanism extracting angular momentum
from the X-ray binaries. To obtain the the stellar wind-loss
rate, we assume that a fraction of the X-ray luminosity of the NS
is converted into the kinetic energy of a wind of the donor star.
I. e.
\begin{equation}
\frac{GM_{\rm d}\dot{M}_{\rm wind}}{R_{\rm d}}=L_{\rm
X}f_{\Omega}f_{\epsilon},
\end{equation}
where $f_{\Omega}$ is the geometric factor of the X-ray flux
intercepted by the donor star, and $f_{\epsilon}$ is the wind
driving energy efficiency factor (Justham et al. 2006). The X-ray
luminosity of the accreting NS can be written as
\begin{equation}
L_{\rm X}=0.1\dot{M}_{\rm acc}c^{2},
\end{equation}
where $\dot{M}_{\rm acc}$ is the accretion rate of the NS. By
equations (3), (4), (5), and Kepler's third law, we obtain (see
also equation (15) in Justham et al. 2006)
\begin{equation}
\dot{J}_{\rm MB}={-B}_{\rm d}\left(\frac{{\psi}{\dot{M}_{\rm
acc}}{M}}{a^3}\right)^{1/2} {\left(\frac{{R_{\rm
d}}^{15}}{G{M_{\rm d}}^3}\right)}^{1/4},
\end{equation}
where $\psi=0.1f_{\Omega}f_{\epsilon}c^{\rm 2}$ is a synthetic
parameter. Since $\dot{J}_{\rm MB}\propto B_{\rm d}\psi^{1/2}$,
the calculated results are more sensitive to $B_{\rm d}$ than to
$\psi$. Therefore, we only consider the influence of magnetic
field on the calculated results. Similarly to \cite{just06}, we
take $f_{\Omega}= 0.01$, $f_{\epsilon}= 10^{-3}$, so
$\psi={10}^{\rm {-6}}c^{\rm 2}$. The angular momentum loss
rate was found to decrease when the donor star becomes completely
convective \citep{rapp83,spru83}. Therefore, in the stellar code
we stop the magnetic braking mechanism when the donor star mass is
less than $0.3~\rm M_{\odot}$, at \textbf{the} point where the star enters
the fully convective stage.

\subsection{Spin and magnetic field evolution of the NS}
In the stellar evolution code, we also take into account the
evolution of the spin and the surface magnetic field ($B_{\rm
NS}$) of NSs. For the spin evolution, the NS is assumed to
experience three evolutionary stages comprising an accretion
phase, a propeller phase, and a radio phase (see also section 2.3
of Liu \& Chen 2011 and Table 1 of Chen \& Liu 2013). In addition,
there also exists a spin-down torque originating from the magnetic
coupling between the magnetic field of the spinning NS and the
outer region of the accretion disc \citep{gho79,rude89b}.
Considering the accretion torque, the propeller torque, and the
spin-down torque caused by the interaction between the magnetic
field and the accretion disc, the total torque exerted on the NS
can be written as \citep{dai06}

\begin{equation}
N =
\left\{
\begin{array}{l}
-\dot{M}_{\rm d}\sqrt{GM_{\rm
NS}R_{0}}\left[\xi(1-\omega)+\frac{\sqrt{2}}{3}
(1-2\omega+\frac{2\omega^{2}}{3})\right],
 \\ (\omega\leq1) \\
\\
-\dot{M}_{\rm d}\sqrt{GM_{\rm
NS}R_{0}}\left[\xi(1-\omega)+\frac{\sqrt{2}}{3}
(\frac{2}{3\omega}-1)\right],\\ (\omega>1) \\
\end{array}
\right.
\end{equation}
where $N={\rm d}J_{\rm NS}/{\rm d}t=I{\rm d}\Omega_{\rm NS}/{\rm
d}t$, $R_{0}$ is the magnetospheric radius of the NS, $\xi$ is a
parameter depending on the structure of the magnetosphere. In
calculation, we take a moderate value $\xi$ = 0.5, and a constant
moment of inertia $I=10^{45}~\rm g\,cm^{2}$. The fastness
parameter $\omega=\Omega_{\rm NS}/\Omega_{0}$, where $\Omega_{\rm
NS}$ is the angular velocity of the NS, $\Omega_{0}$ is the
(Keplerian) angular velocity at the magnetosphere radius.

For the magnetic field evolution, we simply take a
phenomenological form of magnetic field decay due to the accretion
given by \cite{shib89}. I. e.
\begin{equation}
B_{\rm NS}=\frac{B_{\rm i}}{1+\triangle M/10^{-4} \rm M_{\odot}},
\end{equation}
where $B_{\rm i}$ and $\triangle M$ are the initial magnetic field
and the accreted mass of the NS, respectively.

In our calculation, we assume that the NS has passed through the
death line before it was recycled. Observations show that the
critical voltage (generated above the polar cap of the pulsar)
line satisfies $B_{\rm NS}/P_{\rm s}^{2}\approx2\times10^{11}$
($P_{\rm s}$ is the pulse period, Rawley et al. 1986), and a
theoretical study also presented a similar estimate by
\cite{rude75}.  In addition, the pulse periods of known radio
pulsars are in the range of 1.4 ms -- 11 s \citep{man04}. The
final spin period of the NS during the recycled process is
insensitive to its initial spin period and magnetic field
\citep{wang11}. Therefore, we take an initial spin period of 10
s, a typical magnetic field of $10^{12}$ G, and a canonical
initial NS mass of 1.4 $\rm M_{\odot}$.

\section{Simulated results}
\begin{figure}
 \includegraphics[width=0.5\textwidth]{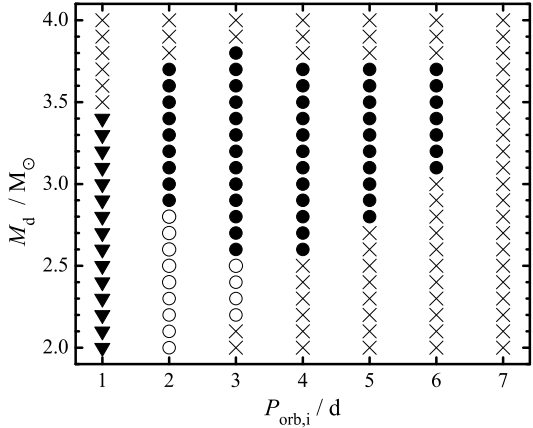}
  \caption{Distribution of the initial Ap/Bp star masses $M_{\rm {d,
i}}$ and the initial orbital periods $P_{\rm {orb, i}}$ of IMXBs
that can evolve into binary pulsars by the anomalous magnetic
braking of Ap/Bp stars when $B_{\rm d} $ = 1000 G. The filled
circles and open circles represent the IMBPs, and LMBPs,
respectively. The crosses correspond to IMXBs that experience
dynamically unstable mass transfer and common envelope evolution.
The filled triangles denote systems where the mass transfer
persists after a Hubble time.}
 \label{fig:fits}
\end{figure}

Fig. 1 summarizes our simulated results by showing the
evolutionary fate of IMXBs in the initial Ap/Bp star masses vs.
the initial orbital periods diagram. It is seen that, if the
initial orbital period in the range of 2 -- 4 d, IMXBs with a
relatively low-mass ($2.0 - 2.8~\rm M_{\odot}$ ) Ap/Bp stars can
evolve into LMBPs with a He WD. While for relatively high-mass
Ap/Bp stars with a mass of $2.6 - 3.8~\rm M_{\odot}$, IMXBs can
form IMBPs with a CO WD if the initial orbital period is in the
range of 2 -- 6 d, depending on the donor star mass. If the donor
star is in a close binary ($P_{\rm orb}$ less than 1 d), during
the Roche-lobe overflow the orbital separation continuously
decreases because the material is transferred from the more
massive donor star on to the less massive NS. The shrinking
Roche-lobe causes the donor star to overfill its Roche-lobe even
more, unstable mass transfer commences and the system experiences
common envelope evolution (Tauris et al. 2000). Similarly, if the
initial orbital period is more than 7 d, long timescale nuclear
evolution of the donor star before the mass transfer causes the
formation of a deep convective envelope. Subsequently, a runaway
material transfer is triggered, giving rise to the spiral-in of
the NS and common envelope evolution. The anomalous magnetic
braking mechanism is very efficient in removing angular momentum
from binaries. Employing the magnetic braking, IMXBs with a short
orbital period ($\la 1$ d) evolve into relatively compact X-ray
binaries until the donor star mass is less than $0.3~\rm
M_{\odot}$. Subsequently, gravitational radiation causes the
continuous shrinking of \textbf{the} Roche lobe, and triggers a
new mass-transfer episode. Therefore, in these systems the mass
transfer persists after a Hubble time (see the filled triangles in
Fig. 1 ). Finally, the donor star can not evolve into WD, but
remains with a semi-degenerate He core. It is worth noting that,
considering anomalous magnetic braking in our work, the initial
parameter space that can form binary pulsars is smaller than found
in previous works (Tauris et al. 2000; Shao \& Li 2012).

In order to illustrate the influence of the magnetic field on the
orbital period of IMBPs, in Fig. 2 we present the correlation
between the final orbital period $P_{\rm {orb,f}}$ and the surface
magnetic field $B_{\rm d}$ of Ap/Bp stars with different masses
when the initial orbital period of IMXBs $P_{\rm {orb,i}}$ = 2.0
d. For the same donor star, the simulated results of four
different magnetic field including 0, 200, 1000, and 3000 G are
shown. The filled squares, the open circles and the open squares
correspond to the initial Ap/Bp star with mass 2.5 $\rm
M_{\odot}$, 3.0 $\rm M_{\odot}$ and 3.5 $\rm M_{\odot}$,
respectively. As shown in this figure, when $B_{\rm d}$ = 0 IMXBs
evolve into IMBPs with a relatively long orbital period (10 -- 50
d). However, relatively weak magnetic braking of Ap/Bp stars
($B_{\rm d}$ = 200 G) can result in a moderately long orbital
period (3 -- 8 d). And strong anomalous magnetic braking of Ap/Bp
stars ($B_{\rm d}$ = 1000, or 3000 G) can induce the birth of
compact IMBPs with an orbital period of 0.2 -- 1 d when the donor
star mass $M_{\rm d}\geq 3~\rm M_{\odot}$.

\begin{figure}
 \includegraphics[width=0.5\textwidth]{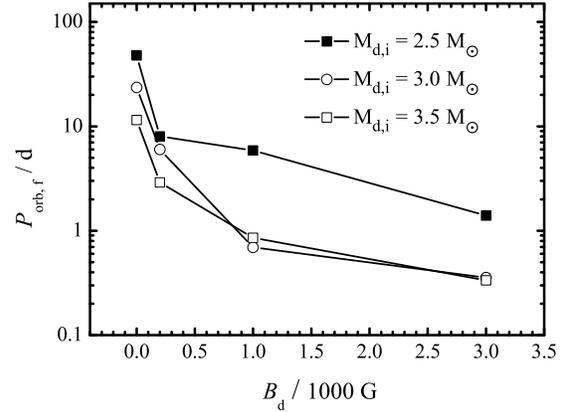}
  \caption{ The final orbital periods $P_{\rm {orb,f}}$ of IMBPS as
a function of the surface magnetic fields $B_{\rm d}$ of Ap/Bp
stars when the initial orbital period of IMXBs $P_{\rm {orb,i}}$ =
2.0 day. The solid squares, open circles, and open squares
correspond to the donor star masses of 2.5, 3.0, and 3.5
$M_{\odot}$, respectively.}
 \label{fig:fits}
\end{figure}

\begin{table}
\begin{center}
\caption{ Measured parameters of 18 known IMBPs.\label{tbl-1}}
\begin{tabular}{cccc}
\hline\hline
PSR & $P_{\rm s}$ &  $ P_{\rm orb}$ & References\\
&(ms)&(days)& \\
\hline
long orbital period &IMBPs& ($P_{\rm orb}\ga 3$ d)&\\
 \hline
J1420$-$5625 & 40.3   & 34.1   &1 \\
J1810$-$2005 & 32.8   & 15.01  &2 \\
J1904+0412   & 71.1   & 14.93  &2 \\
J1454$-$5846 & 45.2   & 12.42  &2  \\
J1614$-$2230 & 3.15   & 8.69   & 3,4\\
J0621+1002   & 28.9   & 8.319   &5,6\\
J1022+1001   & 16.5   & 7.805  &6  \\
J2145$-$0750 & 16.1   & 6.839  &7  \\
J1603$-$7202 & 14.8   & 6.309  &8   \\
\hline
short orbital period& IMBPs &($P_{\rm orb}\la 3$ d)&\\
 \hline
J1157$-$5112 & 43.6   & 3.507  &9  \\
J1528$-$3146 & 60.8   & 3.18   &10  \\
J1439$-$5501 & 28.6   & 2.12   & 11,12\\
J1232$-$6501 & 88.3   & 1.86  &2,13\\
J1435$-$6100 & 9.35   & 1.35  &2,13\\
B0655$+$64   & 195.7  & 1.03  &14,15\\
J1802$-$2124 & 12.6   & 0.699  &11,16\\
J1757$-$5322 & 8.87   & 0.453  &17\\
J1952+2630   & 20.7   & 0.392  &18\\
\hline\hline
\end{tabular}
     \item References. (1)\cite{hobb04};
     (2)\cite{cami01}; (3)\cite{craw06}; (4)\cite{demo10}; (5)\cite{spla02}; (6)\cite{cami96}; (7)\cite{bail94}; (8)\cite{lori96};
     (9)\cite{edwa01a}; (10)\cite{jaco07}; (11)\cite{faul04};
     (12)\cite{lori06}; (13)\cite{man01}; (14)\cite{jone88}; (15)\cite{lori95};
     (16)\cite{ferd10}; (17)\cite{edwa01b};
     (18)\cite{knis11}.\\

\end{center}
\end{table}

\begin{figure}
 \includegraphics[width=0.5\textwidth]{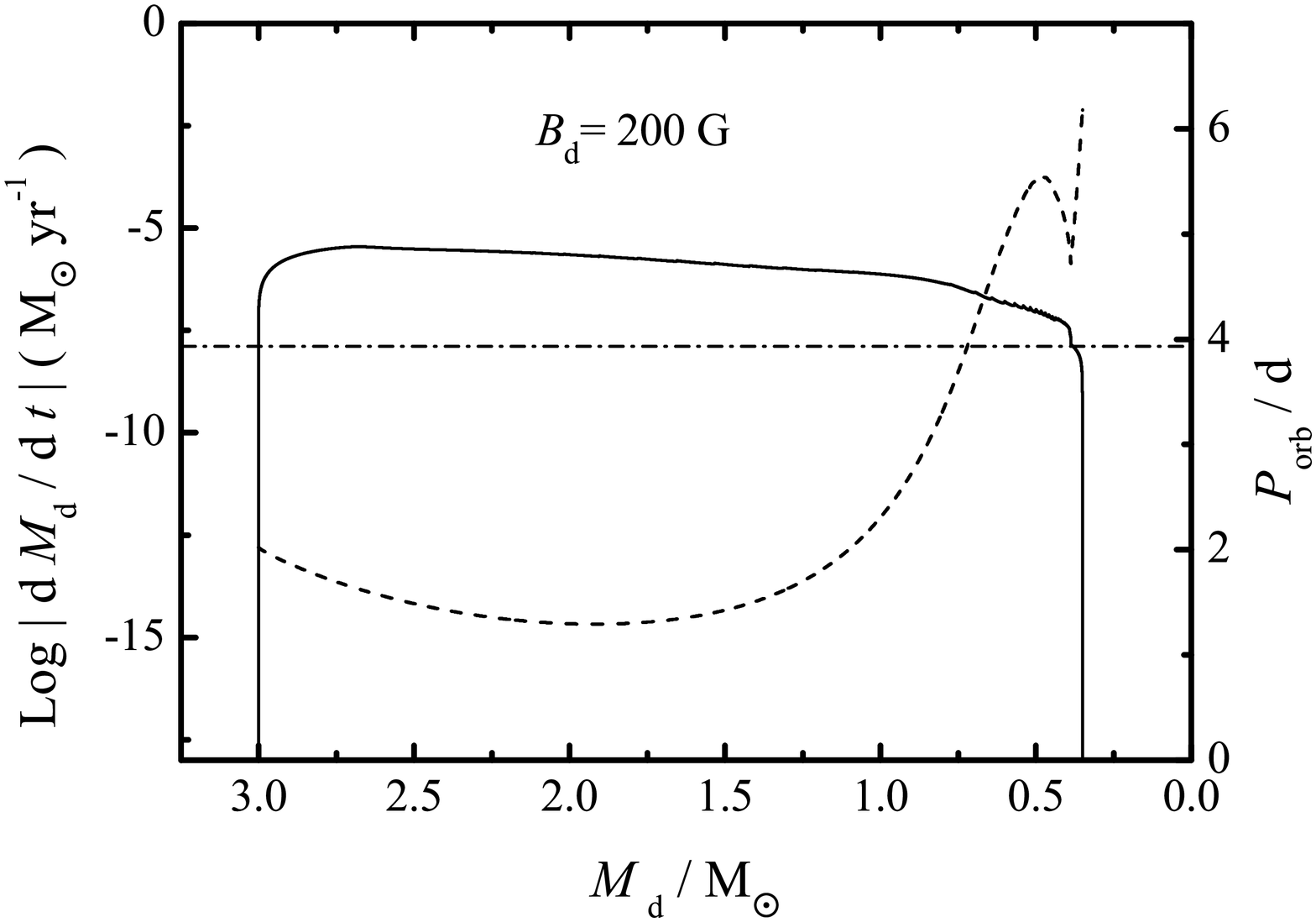}
 \includegraphics[width=0.5\textwidth]{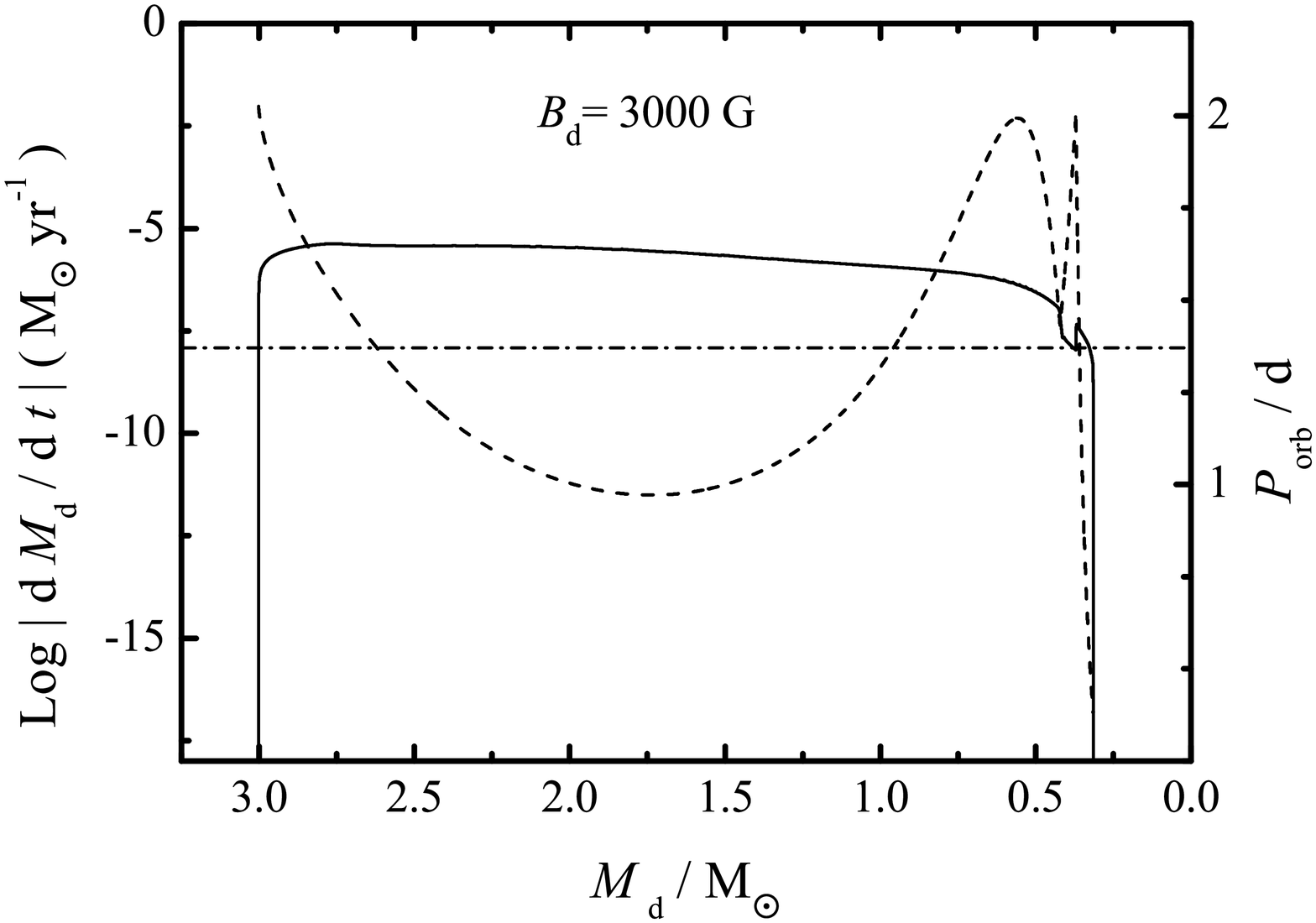}
  \caption{Evolutionary tracks of a NS IMXB with an Ap/Bp star of
$3.0~\rm M_\odot$ and an initial orbital period of $P_{\rm {orb,
i}}$ = 2.0 d. In the upper panel, and lower panel, the donor star
has a surface magnetic field of 200, and 3000 G, respectively. The
solid and dashed curves represent the evolution of the mass
transfer rate and the orbital period, respectively. The horizontal
line corresponds to the Eddington accretion rate.}
 \label{fig:fits}
\end{figure}

In Fig. 3, we plot the evolutionary track of the mass transfer
rate and the orbital period of the NS IMXB with a donor star of
$3.0~\rm M_\odot$ and an initial orbital period of $P_{\rm {orb,
i}}$ = 2.0 d under different magnetic fields, $B_{\rm d}$ = 200
and 3000 G.  Fig. 4 shows the evolution of the NS mass and the
donor star mass when the donor star has magnetic field $B_{\rm d}$
= 200, 3000 G. With the exhaustion of central H, the donor star
starts to fill its Roche-lobe at the age of 322.1 Myr. Because the
material transfer from the more massive donor star to the less
massive NS, the first stage is the thermal timescale mass transfer
and lasts about 1 Myr. The rapid mass transfer occurs at a high
rate of about $10^{-6}~{\rm M_{\odot}yr^{-1}}$ until the donor
star mass decreases to $1.4~M_\odot$. In the first stage, the NS
only accretes about 0.01 $M_\odot$ ($1.5\times 10^{-8}~{\rm
M_{\odot}yr^{-1}}\times 1~ \rm Myr$), and $99.4\%$ of the
transferred material is ejected by the radiation pressure of the
NS. In the second stage, the nuclear burning of remaining H in the
core drove the mass transfer at a rate of $10^{-8}-10^{-7}~{\rm
M_{\odot}yr^{-1}}$. The NS accreted $0.06-0.08~\rm M_\odot$ in a
time-interval of 4 -- 6 Myr. This is consistent with the
conclusion of \cite{pods02}, that IMXBs spent more than $80\%$ of
their X-ray active lifetime as LMXBs. Though the NS only accretes
less than $0.1~M_\odot$ of the transferred material, it can be
sufficiently spun up to 3.57 ms ($B_{\rm d}=200$ G) or 5.51 ms
($B_{\rm d}=3000$ G) (see Fig. 5). If the donor star has a weak
surface magnetic field of 200 G, the orbital period continuously
increases to 5.5 d, then decreases to 4.5 d, and then increases
again to 6.0 d. However, IMXBs including Ap/Bp stars with a strong
magnetic fields (3000 G) can evolve into a compact IMBPs with an
orbital periods of around 0.35 d. Therefore, it seems that strong
magnetic braking of Ap/Bp stars could possibly play an important
role in forming IMBPs with a short orbital period, as an
alternative to post-CE evolution.

\begin{figure}
\centering
\includegraphics[width=0.5\textwidth]{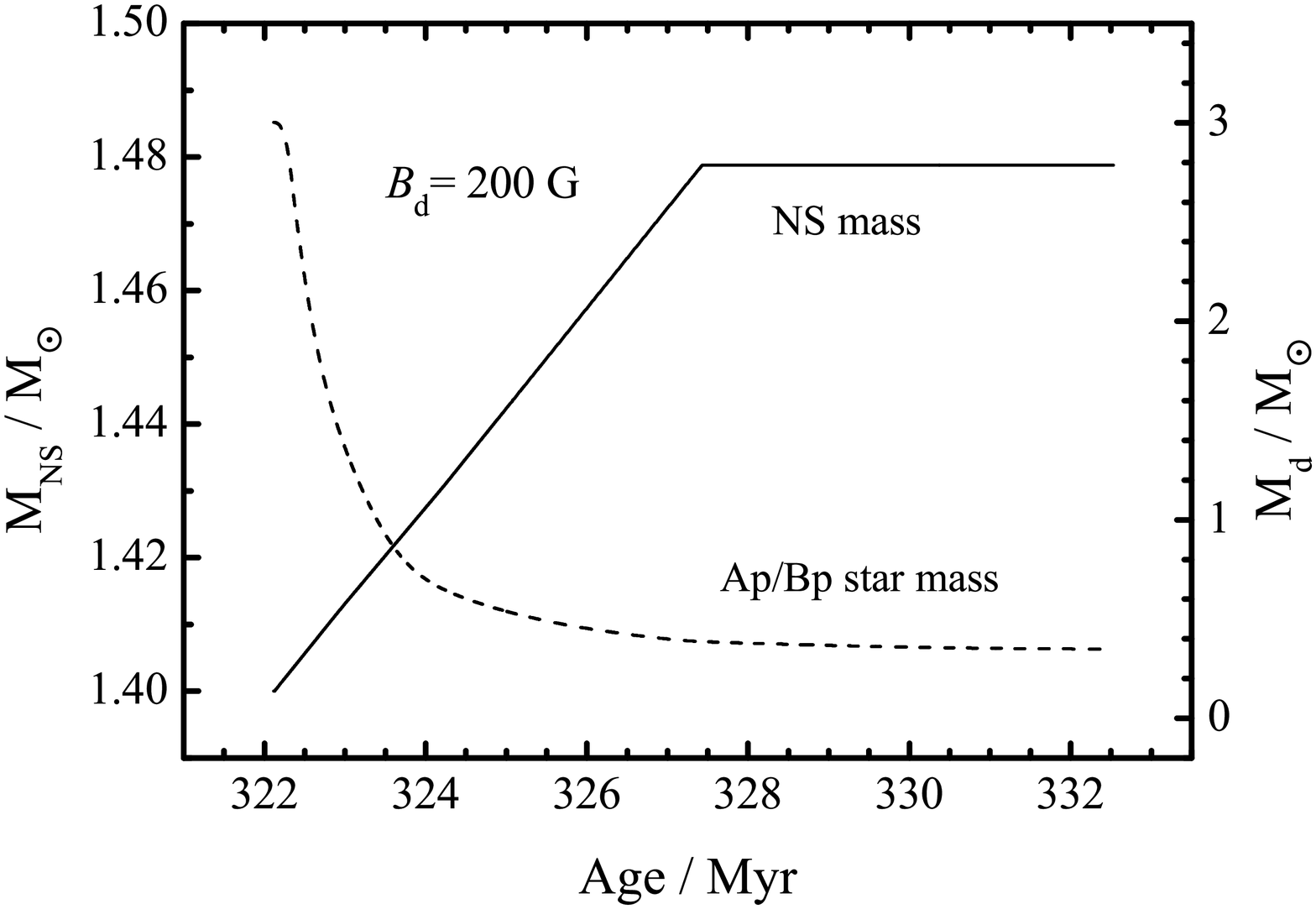}
 \includegraphics[width=0.5\textwidth]{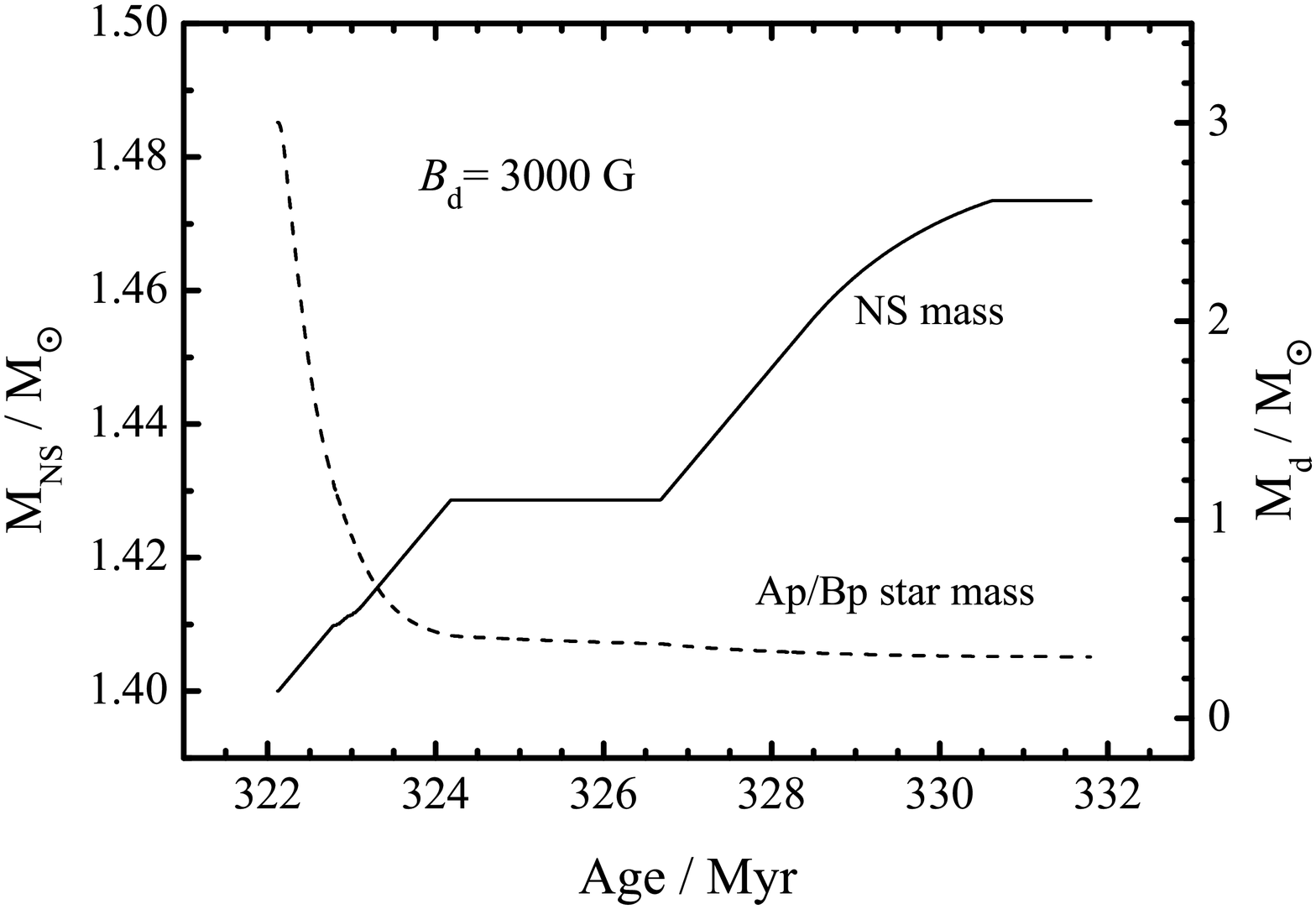}
  \caption{Evolutionary tracks of a NS IMXB with an Ap/Bp star of
$3.0~\rm M_\odot$ and an initial orbital period of $P_{\rm {orb,
i}}$ = 2.0 d. In the upper panel, and lower panel, the donor star
has a surface magnetic field of 200, and 3000 G, respectively. The
solid and dashed curves represent the evolution of the NS mass and
the Ap/Bp star mass, respectively.}
 \label{fig:fits}
\end{figure}

Fig. 6 shows the distribution of our simulated results in the
final orbital period vs. the final spin period plane of IMBPs when
the magnetic field of Ap/Bp stars $B_{\rm d}$ = 1000 G. As shown
in this figure anomalous magnetic braking of Ap/Bp stars can
evolve IMXBs into compact IMBPs with a short orbital period (less
than 10 d) and a small spin period (less than 20 ms). The minimum
of the orbital period of IMBPs forming by this evolutionary
channel is about 0.6 d. In Table 1, we compile the observed
orbital period and spin period for 18 IMBPs. To test our
evolutionary model, 18 IMBPs by the stars are also shown in Fig.
6. One can see that our simulated results can only account for the
formation of a few IMBPs. The far majority of the observed compact
IMBPs tend to have long spin-periods. This discrepancy may
originate from our accretion model, in which we assume the
accretion rate of the NS $\dot{M}_{\rm NS}={\rm min}[\dot{M}_{\rm
Edd}, \mid\dot{M}_{\rm d}\mid]$. Actually, $\dot{M}_{\rm
NS}=f\times{\rm min}[\dot{M}_{\rm Edd}, \mid\dot{M}_{\rm d}\mid]$,
and $f<1$ \citep[see also][]{taur13,laza14}. In Fig.5, we show the
influence of the parameter $f$ on the final spin period of the NS.
One can see that, a moderate parameter $f = 0.5$ can result in a
twice increase of the spin period.

\begin{figure}
\centering
 \includegraphics[width=0.5\textwidth]{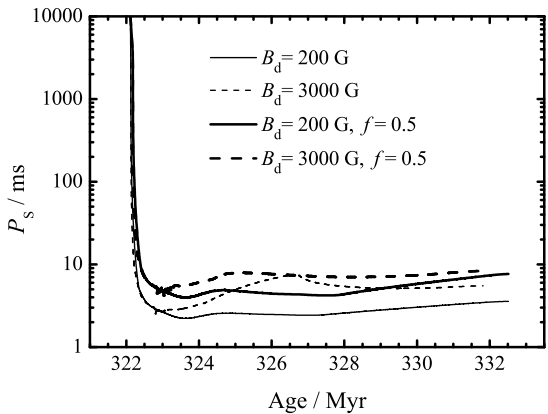}
  \caption{Evolution of spin period of the NS in an IMXB with an
Ap/Bp star of $3.0~\rm M_\odot$ and an initial orbital period of
$P_{\rm {orb, i}}$ = 2.0 d. The solid and dashed curves correspond
the system in which the donor star has a surface magnetic field of
200, and 3000 G, respectively. }
 \label{fig:fits}
\end{figure}

\section{Discussion and summary}
In this paper, we aim at a peculiar binary pulsar problem, namely
the evolutionary origin of compact IMBPs with an orbital period of
less than 3 d. Considering an anomalous magnetic braking model of
Ap/Bp stars proposed by \cite{just06}, we have performed
evolutionary calculations of NS + Ap/Bp star binaries consisting a
$1.4~\rm M_{\odot}$ NS and a 2.0 -- 4.0 $\rm M_{\odot}$ Ap/Bp star
companion, and have tested whether this evolutionary channel can
produce compact IMBPs. Our main results and conclusions are
summarized as follows.

1. Assuming the surface magnetic fields of Ap/Bp stars are 1000 G,
IMXBs with a 2.6 -- 3.8 $\rm M_{\odot}$ donor star and an orbital
period of 2.0 -- 6.0 d can evolve into IMBPs. Beyond this initial
parameter space, most IMXBs would undergo common-envelope
evolution due to the unstable mass transfer process.

2. About $90\%$ of our simulated IMBPs have an orbital period less
than 3\,d. Therefore, we propose that IMXBs including Ap/Bp stars
with a relatively strong magnetic field are potential progenitors
of compact IMBPs, especially in case a NS cannot survive the
spiral-in during a CE.

3. The efficiency of magnetic braking plays a vital role for the
orbital period of IMBPs. If Ap/Bp stars have a weak magnetic field
of 200 G, such an evolutionary channel leads to the formation of
IMBPs with a relatively long orbital period of 3 -- 8 d. However,
for a strong magnetic field of more than 1000 G, compact IMBPs
with an orbital period of less than 1 d can be formed.

4. When $B_{\rm d}=1000$ G, anomalous magnetic braking scenario
only gives rise to the formation of IMBPs with a short spin-period
of less than 20 ms, which is obviously lower than the observed
period of most IMBPs. There may be two cases to address this
discrepancy. In the first case, the NS accretion model in
our calculation should be improved. In the second case,
there exist other evolutionary channels to form IMBPs with a
relatively long spin-period (see also Tauris et al. 2012; Chen \&
Liu 2013; Lazarus et al. 2014). It is very interesting that
anomalous magnetic braking scenarios can form IMBPs with an
orbital period of more than 0.6 d and a spin period of less than
10 ms, which cannot be produced by the NS + He star evolutionary
channel (see also Fig. 4 of Chen \& Liu 2013).

\begin{figure}
\centering
 \includegraphics[width=0.5\textwidth]{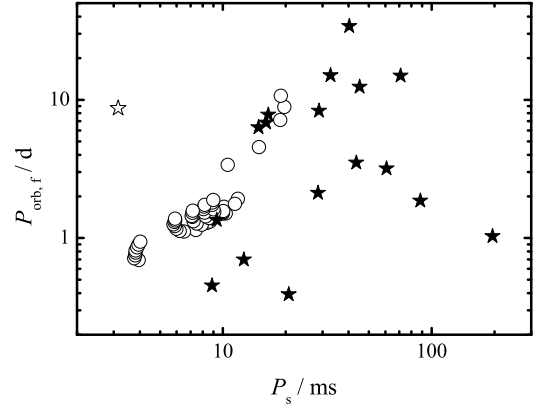}
  \caption{Comparison between our simulated results (open circles)
and the observed data (solid stars) in the final orbital period
vs. the final spin period of IMBPs diagram when the magnetic field
of Ap/Bp stars $B_{\rm d}$ = 1000 G. Note: The open star
represents PSR J1624-2230, which is so far the only known IMBP
formed via Case A Roche-lobe overflow.}
 \label{fig:fits}
\end{figure}

Certainly, there exist many uncertainties in our calculation.
First, for initial input parameters we adopt a canonical NS mass
of 1.4$\rm M_{\odot}$. However, both theoretical studies (see also
Nomoto 1984) and observations (e. g. Kiziltan et al. 2010; Zhang
et al. 2011) argue that the initial masses of NSs have a large
range distribution. Recently, two groups suggested that the
peculiar IMBP PSR J1614--2230 could have evolved from an IMXB with
a heavy NS (about $1.6 - 1.7M_{\odot}$, Lin et al. 2011; Tauris et
al. 2011). Subsequently, \cite{shao12} found that the initial
parameter space (in the initial orbital period vs. the initial
donor star mass diagram) forming binary pulsars expands when the
NS mass increases. Secondly, we simply take a constant magnetic
field strength ($B_{\rm d}=1000$ G). Actually, Ap/Bp stars may
have a large distribution range of 100 -- 10000 G in the surface
magnetic field (Moss 1989; Braithwaite \& Spruit 2004). This would
significantly influence the evolutionary fate of IMXBs. Thirdly,
for the X-ray irradiation-driven stellar wind process by the mass
accretion, in this calculation we take a constant synthetic
parameter $\psi$ like \cite{just06}. However, this parameter may
change from one system to another. Fourthly, the prescription for
magnetic braking is highly uncertain. Fifthly, the magnetic field
decay equation of the NS due to accretion needs to be improved.

A potential caveat of the work presented here is if the tidal
forces can keep the spin of the donor star synchronized with the
orbital motion. Magnetic braking of a star extracting orbital
angular momentum is based on the significant tidal interaction
between the orbit and the star. Studying the spin evolution of the
accreting Algol-type binaries, Dervi\c{s}o\u{g}lu et al. (2010)
found tidal interaction between the components and the orbit is
too weak to efficiently remove spin angular momentum from the
accreting component. The weak energy dissipation for stars with
radiative envelopes should be responsible for this weak tide.
Therefore, magnetic braking of Ap/Bp stars need an efficient
energy dissipation mechanism that is stronger than gravity wave
dissipation (Zahn 2005). In addition, at the end of the
mass-transfer phase the donor star has a relatively thin envelope.
It is unclear if the tidal forces are still efficient for a thin
envelope. This issue goes beyond the scope of this work, and we
plan to explore it in a subsequent work.

\section*{Acknowledgments}
We are grateful to the anonymous referee for helpful comments.
This work was partly supported by the National Science Foundation
of China (under grant number 11173018, U1331117), Innovation
Scientists and Technicians Troop Construction Projects of Henan
Province, China.

\bsp

\label{lastpage}

\end{document}